\documentclass{article}
\usepackage[dvips]{graphicx}

\begin{document}

\title{\textbf{Stable solitons in coupled Ginzburg-Landau equations describing
Bose-Einstein condensates and nonlinear optical waveguides and cavities}}
\author{Hidetsugu Sakaguchi$^{a}$ and Boris A. Malomed$^{b}$ \\
$^{a}$ Department of Applied Science for Electronics and Materials, \\
Interdisciplinary Graduate School of Engineering Sciences, Kyushu
University, \\
Kasuga, Fukuoka 816-8580, Japan\\
$^{b}$Department of Interdisciplinary Studies, Faculty of Engineering,\\
Tel Aviv University, Tel Aviv 69978, Israel}
\maketitle

\begin{center}
\bigskip {\Large Abstract}
\end{center}

We introduce a model of a two-core system, based on an equation of the
Ginzburg-Landau (GL) type, coupled to another GL equation, which may be
linear or nonlinear. One core is active, featuring intrinsic linear gain,
while the other one is lossy. The difference from previously studied models
involving a pair of linearly coupled active and passive cores is that the
stabilization of the system is provided not by a linear diffusion-like term,
but rather by a cubic or quintic dissipative term in the active core.
Physical realizations of the models include systems from nonlinear optics
(semiconductor waveguides or optical cavities), and a double-cigar-shaped
Bose-Einstein condensate with a negative scattering length, in which the
active ``cigar'' is an atom laser. The replacement of the diffusion term by
the nonlinear loss is principally important, as diffusion does not occur in
these physical media, while nonlinear loss is possible. A stability region
for solitary pulses is found in the system's parameter space by means of
direct simulations. One border of the region is also found in an analytical
form by means of a perturbation theory. Moving pulses are studied too. It is
concluded that collisions between them are completely elastic, provided that
the relative velocity is not too small. The pulses withstand multiple
tunneling through potential barriers. Robust quantum-rachet regimes of
motion of the pulse in a time-periodic asymmetric potential are found as
well. \newline

\section{Introduction}

Ginzburg-Landau (GL) equations represent a class of universal mathematical
models which describe pattern formation in various nonlinear media \cite
{AransonKramer}. One of the most fundamental types of patterns are solitary
pulses (SPs), in loose terms often called solitons. In particular, the
simplest generic species of the GL equations, viz., the cubic complex one,
gives rise to a well-known exact SP solution \cite{Lennart}. However, this
solution is unstable (as the equation includes a linear gain term, which
makes the zero solution unstable, precluding stability of any solitary
pattern). Therefore, search for physically relevant models of the GL type
that give rise to stable pulses has attracted much attention. One
possibility is to introduce a cubic-quintic GL equation with linear loss and
cubic gain, nonlinear stability being provided by a quintic loss term.
Stable SPs in equations of the latter type have been studied in detail \cite
{CQ}. Another model, which finds a straightforward physical realization in
terms of dual-core nonlinear optical fibers, was proposed in Ref. \cite
{Winful}. In this system, one core carries linear gain, while the other one
is lossy, the corresponding model being based on a system of two linearly
coupled cubic GL equations (in fact, the one corresponding to the lossy core
may be a linear equation).

Detailed investigations have demonstrated that the latter model supports
stable stationary SPs \cite{Javid,Javid2} and their bound states \cite
{Javid3} in broad parametric regions. Stable moving pulses, randomly
wandering ones, and breathers (both standing and moving) have also been
found in this system \cite{we}. The model was subsequently generalized to
combine it with dispersion-management \cite{DM} and
wavelength-division-multiplexing (i.e., multi-channel) \cite{WDM} schemes,
which opens a way for applications to fiber-optic telecommunications. It was
also shown that a model of the same type can support stable dark solitons 
\cite{dark-soliton}.

In all the cases, the stability of SP solutions in this model was provided
by a linear dispersive-loss term, which accounted for the bandwidth-limited
character of the gain in the active core (formally, that term is tantamount
to diffusion). Without this ingredient, the model can generate only unstable
SPs. However, the natural bandwidth of the optical gain (which is usually
provided for by the Erbium dopant \cite{Er}) is very broad, therefore the
limitation of the gain bandwidth should be enforced by optical filters
specially inserted into the system. A related problem is that, in the case
of other physical systems which may be described by equations of the GL
types, such as optical cavities or planar waveguides and Bose-Einstein
condensates (BECs, see below), the diffusion term is not physically possible
at all.

An alternative way to stabilize SPs is to use a nonlinear dissipative term,
which may be cubic or quintic. In the applications to nonlinear optics, a
cubic loss term is naturally generated by two-photon absorption, which is a
strong effect in semiconductor waveguides and semiconductor-doped glasses
(see, e.g., recent works \cite{semi} and \cite{doped}, respectively, and
references therein). The investigation of this possibility is a subject of
the present work. Actually, the system with nonlinear loss and without
filtering/diffusion is not just another version of the above-mentioned
dual-core fiber-optic model, but it also directly applies to the description
of other physical media, namely, nonlinear optical cavities and BECs, as it
is explained below.

In normalized units, coupled equations of the type described above can be
cast in the form 
\begin{eqnarray}
u_{t} &=&\gamma u+i\gamma _{2}u_{xx}+\left( i\sigma _{1}-\sigma _{2}\right)
|u|^{2}u+iv,  \label{u} \\
v_{t} &=&-\left( \Gamma +i\chi \right) v+i\gamma _{2}v_{xx}+iu,  \label{v}
\end{eqnarray}
where, in the case of the optical systems, $u$ and $v$ are amplitudes of
electromagnetic waves in two cores of the system, the evolutional variable 
$t $ is either time or propagation distance in the optical cavity (depending
on the physical formulation \cite{cavity}), or the propagation distance in
the dual-core optical fiber, and $x$ is the transverse coordinate in the
cavity (or in a planar waveguide), or the reduced time in the application to
the fibers. Further, the term with $\gamma >0$ in Eq. (\ref{u}) accounts for
the gain in the active subsystem, the constant of linear coupling between
the cores is normalized to be $1$, $\Gamma >0$ is the dissipative constant
in the lossy subsystem, $\chi $ is a possible frequency- (if $t$ is time) or
wavenumber- (if $t$ is the propagation distance) mismatch between the cores,
and $\gamma _{2}$, which may be assumed positive, is the
dispersion/diffraction coefficient (in fact, $\gamma _{2}$ may be given a
fixed value by means of obvious rescaling; $\gamma _{2}=5$ will be chosen
below, as this value is convenient to display numerical results). Finally, 
$\sigma _{2}$ and $\sigma _{1}$ are, respectively, coefficients of the Kerr
nonlinearity and nonlinear loss in the active core. It is assumed that, in
most cases, the field in the lossy core is much weaker than in the active
one, therefore nonlinear terms in Eq. (\ref{v}) may be neglected \cite
{Javid2,we}, although properties of SPs in the system including nonlinear
terms in the lossy subsystem are quite similar to those in the system based
on Eqs. (\ref{u}) and (\ref{v}) \cite{Javid,Javid2}. A principal difference
of the model based on Eqs. (\ref{u}) and (\ref{v}) from the previously
considered ones is that the coefficient in front of the term $iu_{xx}$ in
Eqs. (\ref{u}) and (\ref{v}) is real, while the nonlinear coefficient in Eq.
(\ref{u}) is complex; previously, exactly the opposite case\ was considered 
\cite{Javid,Javid2,we}.

In the application to BECs, each equation (\ref{u}) and (\ref{v}) may be
realized as a one-dimensional Gross-Pitaevskii equation for a condensate in
a cigar-shaped trap, with the linear coupling induced by tunneling between
them (see, e.g., Ref. \cite{Inguscio}). The second derivatives in Eqs. (\ref
{u}) and (\ref{v}) are then the kinetic-energy terms, with $\gamma
_{2}=\hbar /2m$, where $m$ is the atomic mass. The linear-loss term in Eq. 
(\ref{v}) accounts for evaporation of condensate atoms in the second trap,
while the linear gain in Eq. (\ref{u}) assumes that the corresponding trap
is arranged as an \textit{atom-wave laser} \cite{laser}. Further, the
coefficient $\sigma _{1}$ in Eq. (\ref{u}) is proportional to the scattering
length of atomic collisions in the BEC gas, and $\sigma _{2}>0$ accounts for
effective loss due to two-body collisions \cite{two-body-loss}.

In fact, in many cases a dominant contribution to the nonlinear loss in BECs
is due to three-body collisions \cite{three-body-loss}, hence the
corresponding dissipative term is quintic, and the accordingly modified
equation (\ref{u}) takes the form (again, with $\sigma _{2}>0$) 
\begin{equation}
u_{t}=\gamma u+i\gamma _{2}u_{xx}+\left( i\sigma _{1}-\sigma
_{2}|u|^{2}\right) |u|^{2}u+iv.  \label{quintic}
\end{equation}
Below, we will consider two systems, (\ref{u}), (\ref{v}) and 
(\ref{quintic}), (\ref{v}) with the cubic and quintic nonlinear-loss 
terms, respectively.

The rest of the paper is organized as follows. In Section 2 we present
analytical results for SPs obtained by means of a perturbation theory.
Numerical results are displayed in Section 3, where a parametric region for
the existence of stable SPs is identified. In Section 4, we consider moving
solitons, demonstrating that they collide elastically with each other. We
also consider a generalization of the model including an $x$-dependent
potential, which is relevant for BECs, and may be relevant for the case of
optical planar waveguides too. In that case, we find that moving SPs can
coherently tunnel (many times) through a potential barrier. The paper is
concluded by Section 5.

\section{Analytical results}

Following the lines of Ref. \cite{Winful}, it is possible to develop
analysis of SP solutions, considering them as weakly perturbed nonlinear
Schr\"{o}dinger (NLS) solitons. The linear coupling between the two
equations, as well as the gain and loss terms, are treated as perturbations,
which are ordered so that the conservative one, i.e., the linear coupling,
is assumed to be a larger perturbation, while the nonconservative terms are
assumed to be a smaller perturbation. Thus, in the lowest-order
approximation, that ignores the loss and gain but takes the coupling into
regard at the first order of the perturbation theory, the SP has the form 
\begin{equation}
u=\frac{\eta }{\sqrt{\sigma _{1}}}\,\mathrm{sech}\left( \frac{\eta \,x}
{\sqrt{2\gamma _{2}}}\right) \exp \left( \frac{i}{2}\eta ^{2}t\right)
,\,v=V(x)\exp \left( \frac{i}{2}\eta ^{2}t\right) ,  \label{soliton}
\end{equation}
where it is assumed that $\sigma _{1}$ is positive, $\eta $ is an arbitrary
real constant, which is an intrinsic parameter of the soliton family, and
the real function $V(x)$ is a solution to the linear inhomogeneous equation, 
\begin{equation}
\gamma _{2}\frac{d^{2}V}{dx^{2}}-\left( \frac{1}{2}\eta ^{2}+\chi \right) V=-
\frac{\eta }{\sqrt{\sigma _{1}}}\,\mathrm{sech}\left( \frac{\eta \,x}{\sqrt{
2\gamma _{2}}}\right) .  \label{V}
\end{equation}

Note that, in this approximation, the linearly coupled equations conserve
three dynamical invariants: the Hamiltonian and momentum, expressions for
which will not be used here, and the norm $N$, which has the physical
meaning of energy or power in the applications to optics (depending on the
particular interpretation -- it is energy in the fiber, or power in the
cavity or planar waveguide), or the total number of atoms in the case of
BECs, 
\begin{equation}
N=\int_{-\infty }^{+\infty }\left[ \left| u(x)\right| ^{2}+\left|
v(x)\right| ^{2}\right] dx.  \label{N}
\end{equation}

In the next approximation, when the nonconservative perturbations are taken
into account, we assume that $\eta $ may be a slowly varying function of
time [then, the expression $(1/2)\eta ^{2}t$ for the SP phase in Eq. (\ref
{soliton}) is replaced by $(1/2)\int \eta ^{2}(t)dt$]. An evolution equation
for $\eta (t)$ can be derived from the balance equation for $N$ . In the
case of the cubic loss term, which corresponds to Eq. (\ref{u}), it is 
\begin{equation}
\frac{dN}{dt}=2\int_{-\infty }^{+\infty }\left[ \gamma \left| u(x)\right|
^{2}-\Gamma \left| v(x)\right| ^{2}\right] dx-2\sigma _{2}\int_{-\infty
}^{+\infty }\left| u(x)\right| ^{4}dx.  \label{balance}
\end{equation}
In the case of the quintic loss term, corresponding to Eq. (\ref{quintic}), 
$\left| u(x)\right| ^{4}$ in the last term of Eq. (\ref{balance}) is replaced
by $\left| u(x)\right| ^{6}$.

By itself, Eq. (\ref{balance}) is an exact one. To derive the evolution
equation for $\eta (t)$ from it, we substitute the approximation (\ref
{soliton}) for $u$, and make use of Eq. (\ref{V}) (which is solved by means
of the Fourier transform). Then, after straightforward calculations
following the pattern of Ref. \cite{Winful}, the resultant evolution
equation can be obtained in an explicit form if $\chi =0$. With the cubic
loss term, it takes the form 
\begin{equation}
\frac{d\eta }{dt}=2\gamma \eta -\frac{4\sigma _{2}}{3\sigma _{1}}\eta
^{3}-C\Gamma \eta ^{-3},  \label{evolution}
\end{equation}
where $C\equiv \left( \pi ^{2}/6\right) +\zeta (3)\approx 2.845$, and $\zeta 
$ is the Riemann's zeta-function. If the loss term is quintic, Eq. (\ref
{evolution}) is replaced by 
\begin{equation}
\frac{d\eta }{dt}=2\gamma \eta -\frac{16\sigma _{2}}{15\sigma _{1}}\eta
^{5}-C\Gamma \eta ^{-3}.  \label{evolution'}
\end{equation}

Stationary SP solutions are selected from the continuous soliton family as
fixed points (FPs) of Eq. (\ref{evolution}) or (\ref{evolution'}), i.e., as
roots of the expression on the right-hand side of the equation. The roots
can be found in a simple form in the case of Eq. (\ref{evolution'}): 
\begin{equation}
\eta _{\mathrm{FP}}^{4}=\left( 16\sigma _{2}\right) ^{-1}\left[ 15\sigma
_{1}^2\gamma \pm \sqrt{15\sigma_1^2\left( 15\sigma _{1}^2\gamma ^{2}-16C\sigma
_{2}\Gamma \right) }\right] .  \label{FP}
\end{equation}

An elementary consideration demonstrates that, in either model, there may
exist two physical FPs or none, depending on values of the parameters: in
the case of Eq. (\ref{evolution}), physical solutions exist if 
\begin{equation}
2\sigma _{1}^{2}\gamma ^{3}\geq 3C\sigma _{2}^{2}\Gamma ,  \label{analytical}
\end{equation}
and in the case of Eq. (\ref{evolution'}), this condition is replaced by [as
it immediately follows from Eq. (\ref{FP})] 
\begin{equation}
15\sigma _{1}^2\gamma ^{2}\geq 16C\sigma _{2}\Gamma ;  \label{analytical'}
\end{equation}
note that $\gamma _{2}$ does not appear in Eqs. (\ref{analytical}) and (\ref
{analytical'}). Further, it follows from Eqs. (\ref{evolution}) and (\ref
{evolution'}) that the FP corresponding to a smaller value of $\eta $ [for
instance, the one with the lower sign in Eq. (\ref{FP})] is unstable, while
the FP corresponding to larger $\eta $ is stable (within the framework of
the present approximation). In the next section, these analytical
predictions will be compared with results of direct numerical simulations.

Lastly, we notice that the stability of the zero solution is a necessary
condition for the full stability of any solitary pattern. In turn, simple
necessary (but, generally speaking, not sufficient) conditions for the
zero-solution stability take a simple form in the case $\chi =0$ \cite
{Winful}: $\gamma <\Gamma <1/\gamma $ (which implies that $\gamma $ must be
smaller than $1$).

\section{Numerical results}

\begin{figure}[htb]
\begin{center}
\includegraphics[width=12cm]{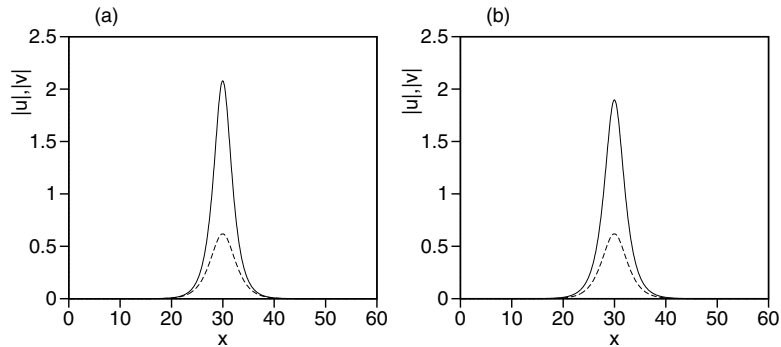}
\end{center}
\caption{Examples of stable stationary solitary-pulse solutions to (a) Eqs. 
(\ref{u}), (\ref{v}), and (b) Eqs. (\ref{quintic}), (\ref{v}), found for (a) 
$\protect\gamma =0.4,\protect\sigma _{1}=1,\protect\sigma _{2}=0.1,
\gamma _{2}=5,\Gamma =1,\protect\chi =0$, and (b) $\gamma
=0.5,\Gamma =1,\protect\gamma _{2}=5,\protect\sigma _{1}=1,\protect\sigma
_{2}=0.05,\protect\chi =0$. The continuous and dashed curves show $|u(x)|$
and $|v(x)|$, respectively.}
\label{fig:1}
\end{figure}

\begin{figure}[htb]
\begin{center}
\includegraphics[width=13cm]{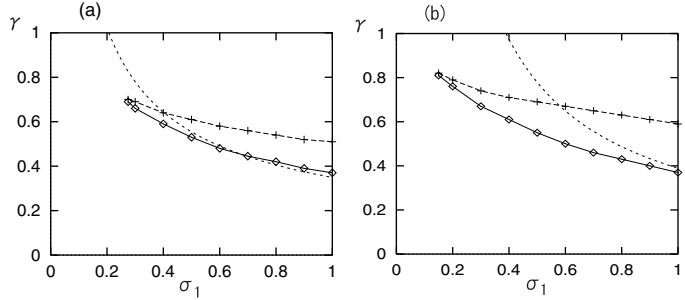}
\end{center}
\caption{A generic example of the stability region for solitary pulses in
the model (\ref{u}), (\ref{v}) (a) and (\ref{quintic}), (\ref{v}) (b), as
found in a numerical form. The fixed parameters are $\protect\sigma _{2}=0.1,
\protect\gamma _{2}=5,\Gamma =1,\protect\chi =0$ in (a), and 
$\sigma_{2}=0.05,\protect\gamma _{2}=5,\Gamma =1,\protect\chi =0$ in (b). 
The curve
denoted by crosses is a border above which some perturbations with high
wavenumbers grow and SPs are destroyed. The curve denoted by diamonds is a
border below which SPs decays to zero. The dashed curve in each panel is the
existence border for the pulses predicted in the analytical form by (a)
Eq.~(11) or (b) Eq.~(12).}
\label{fig:2}
\end{figure}

\begin{figure}[htb]
\begin{center}
\includegraphics[width=6cm]{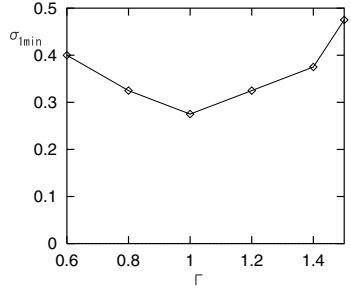}
\end{center}
\caption{The minimum value of the nonlinear coefficient 
$\sigma _{1}$, up to which stable solitary pulses were 
found in Eqs. (\ref{u}), 
(\ref{v}), varying the gain $\protect\gamma $, as 
a function of the linear-loss
coefficient $\Gamma $. The other coefficients were fixed: 
$\sigma_{2}=0.1,\;\protect\gamma _{2}= 5$, and $\protect\chi =0$.}
\label{fig:3}
\end{figure}

\begin{figure}[htb]
\begin{center}
\includegraphics[width=8cm]{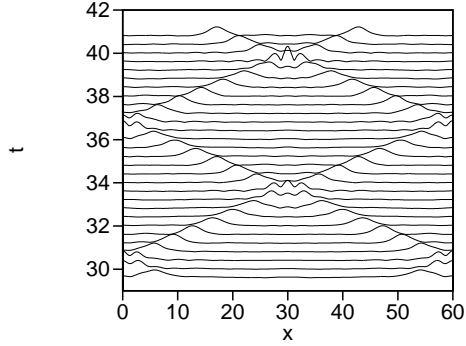}
\end{center}
\caption{A typical example of multiple collisions between two moving
solitons in Eqs. (\ref{u}) and (\ref{v}) with periodic boundary conditions.
The parameters are $\protect\sigma _{2}=0.1,\protect\gamma _{2}=5,\Gamma =1,
\protect\chi =0$ and the spatial period is $L=60$. The initial wavenumbers
of the pulses [see Eqs. (\ref{u}) and (\ref{v})] are $k=\pm 0.5$.}
\label{fig:4}
\end{figure}

\begin{figure}[htb]
\begin{center}
\includegraphics[width=8cm]{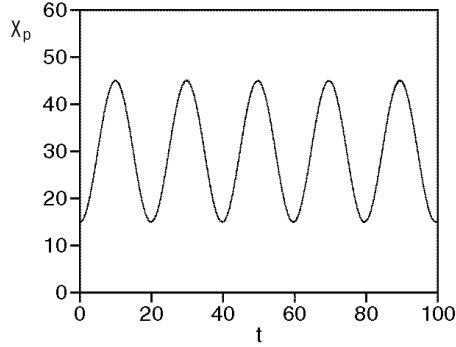}
\end{center}
\caption{Time evolution of the peak position of the oscillating pulse in the
harmonic potential $U(x)=(K/2)(x-L/2)^{2}$ with $K=0.01$ for $\protect\gamma
=0.4,\;\protect\sigma _{2}=0.1,\;\hbar=1,\;m=0.1,\Gamma =1,\protect\chi =0$.}
\label{fig:5}
\end{figure}

\begin{figure}[htb]
\begin{center}
\includegraphics[width=8cm]{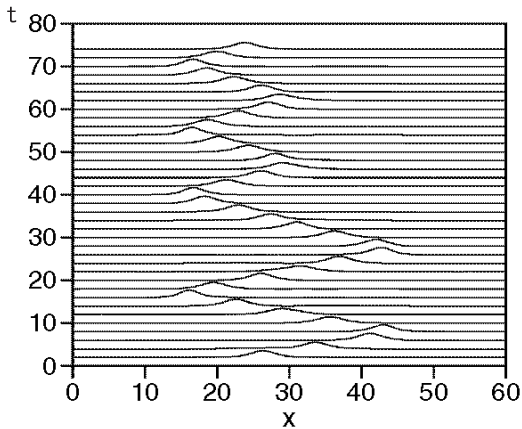}
\end{center}
\caption{An example of multiple coherent tunneling of a traveling solitary
pulse across a narrow potential barrier (19) in the model 
(\ref{u2}), (\ref{v2}). The parameters are $\gamma =0.4,\;
\sigma_{2}=0.1,\;\hbar=1,\;m=0.1,\Gamma =1,\;\chi =0$, and the initial
wavenumber pushing the pulse is $k=0.4$.}
\label{fig:6}
\end{figure}

\begin{figure}[htb]
\begin{center}
\includegraphics[width=12cm]{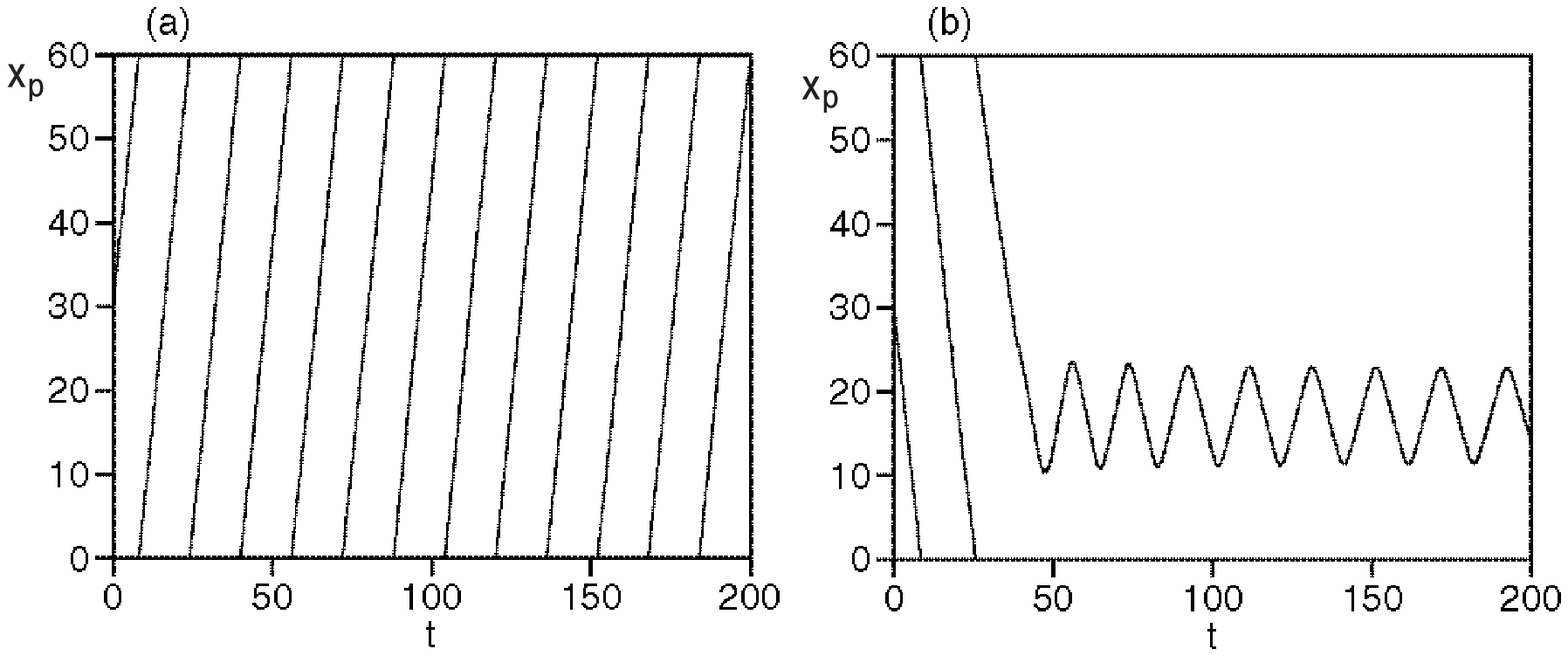}
\end{center}
\caption{An example of the quantum-ratchet effect for traveling pulses in
the time-periodic sawtooth potential (20) for $\protect\gamma =0.4,\;
\sigma _{2}=0.1,\;\hbar=1,\;m=0.1,\Gamma =1,\protect\chi =0$. The initial
velocity is $4$ in (a) and $-4$ in (b). The temporal dependences of the peak
position $X_{p}$ are displayed in both panels.}
\label{fig:7}
\end{figure}

In the numerical investigation, SP solutions were looked for as solutions to
the stationary version of Eqs. (\ref{u}), (\ref{v}) or (\ref{quintic}), (\ref
{v}); then, stability of the obtained solutions was tested in simulations of
the full time-dependent equations. Typical examples of thus found stable SPs
in both models are displayed in Fig. 1.

Results of systematic numerical simulations of Eqs. (\ref{u}), (\ref{v}) and
(\ref{quintic}), (\ref{v}) are summarized in Fig. 2, which displays generic
examples of stability regions for SPs in the parameter plane $\left( \sigma
_{1},\gamma \right) $. These two parameters are chosen for the variation,
while the others are fixed, as they can be readily varied in the
experimental studies of BECs: the gain $\gamma $ is controlled by intrinsic
parameters of the atom laser, and the nonlinearity $\sigma _{1}$ can be
changed by means of the Feshbach resonance \cite{Feshbach}. In the
applications to optics, the gain parameter is also easy to vary.

The analytical predictions (\ref{analytical}) and (\ref{analytical'}) for
the existence of SPs correlate with lower borders of the numerically found
stability regions in Figs. 2(a) and 2(b). As is seen, the perturbative
result is accurate enough for relatively large values of $\sigma _{1}$ and
small values of $\gamma $, when the gain and loss may indeed be regarded as
small perturbations; with the decrease of $\sigma _{1}$ and increase of 
$\gamma $, the perturbations are no longer small, which explains discrepancy
between the analytically predicted and numerically found lower borders in
Figs. 2(a) and 2(b) in this case.

Below the existence border, the perturbation theory predicts that no
stationary SP exists; in accord with this, the numerical simulations show
that, beneath the lower border, any initial pulse decays to zero. On the
other hand, the upper border of the stability regions in Fig. 2 actually
bounds not the existence, but rather stability of the SPs. Above the upper
border, perturbations with high wavenumbers grow and destroy the SPs.

The unperturbed NLS equation supports (bright) solitons only in the case 
\begin{equation}
\gamma _{2}\sigma _{1}>0,  \label{gammasigma}
\end{equation}
see Eqs. (\ref{u}) and (\ref{quintic}). In terms of nonlinear optics, this
condition corresponds to a combination of spatial diffraction or anomalous
temporal dispersion and self-focusing nonlinearity, or normal temporal
dispersion and self-defocusing nonlinearity; in terms if BECs, it
corresponds to the case of negative scattering length \cite{Feshbach}.

An issue of considerable interest is whether the addition of dissipative
terms makes it possible to relax the condition (\ref{gammasigma}). Note that
the exact SP solution to the cubic complex GL equation exists irrespective
of this condition \cite{Lennart}, but that solution is always unstable. In
Ref. \cite{Javid2}, \emph{stable} SPs were found, in the two-core model
including the filtering term, for \emph{both} signs of the product 
$\gamma_{2}\sigma _{1}$. However, our result for the present model, in which 
the filtering term is replaced by the cubic or quintic loss, is that the
condition (\ref{gammasigma}) remains necessary for the existence of SPs. In
fact, fixing $\gamma_{2}$ to be a positive constant, it is interesting to
find the smallest value of the nonlinearity coefficient $\sigma_{1}$ up to
which the stable pulse persists. The result [obtained for Eqs. (\ref{u}) and
(\ref{v}), with the cubic loss term] is shown in Fig. 3, which displays the
smallest SP-supporting value of $\sigma _{1}$ versus the linear-loss factor 
$\Gamma $, as found by varying the linear gain $\gamma $, while $\sigma
_{2},\gamma _{2}$ and $\chi =0$ were fixed.

Similar analysis was performed for models differing from the ones considered
above by adding to Eq. (\ref{v}) the conservative cubic term $i\sigma
_{1}|v|^{2}v$, i.e., essentially the same one as in Eq. (\ref{u}) or (\ref
{quintic}). The result is that the stability regions for SPs are similar to,
although somewhat smaller than, those shown in Fig. 2.

\section{Moving pulses and coherent tunneling}

One of principal distinctions between the models with the nonlinear loss and
ones with the linear diffusion (filtering) is that the models considered in
the present work share the Galilean invariance with the NLS equation. Due to
this reason, solutions for SPs moving at an arbitrary velocity can be
generated by the Galilean transform from any quiescent SP. The moving pulse
solution with an arbitrary wavenumber $k$ can be expressed as 
\begin{equation}
u_{k}=u_0(x-ct,t)e^{ikx-i\omega t},\;\;\;v_{k}=v_0(x-ct,t)e^{ikx-i\omega t},
\label{k}
\end{equation}
where $u_0(x,t)$ and $v_0(x,t)$ represent a stationary-SP solution to Eqs. 
(\ref{u}) and (\ref{v}), and 
\begin{equation}
c=2\gamma _{2}k,\omega =\gamma _{2}k^{2}.  \label{c,omega}
\end{equation}
In the application to BECs, where $\gamma _{2}=\hbar /2m$, Eqs. (\ref
{c,omega}) are tantamount to the usual relations for the momentum and
kinetic energy of a quantum particle, 
\begin{equation}
\hbar k=mc\equiv P,\,\hbar \omega =(\hbar k)^{2}/(2m)\equiv E_{\mathrm{kin}}.
\label{E}
\end{equation}

The availability of the moving pulses suggests numerical experiments aimed
at simulation of collisions between them. A typical example of the
collisions is displayed in Fig. 4, that presents results of simulations of
Eqs. (\ref{u}) and (\ref{v}) in a domain with periodic boundary conditions,
which gives rise to periodic recurrence of the collision. It is obvious from
this figure, and can be observed as a generic situation, unless the relative
velocity of the colliding solitons is too small, that the collisions are,
practically, completely elastic (if the relative velocity is too small, both
SPs decay to zero state after the collision). The same is true for the model
with the quintic loss.

In the case of BECs, an external potential, such as a harmonic magnetic
trap, is usually an important ingredient of the model (in the optical model
describing the spatial evolution of the fields in coupled planar waveguides,
a similar term may describe a spatially modulated profile of the refractive
index in the waveguides). Equations (1) and (2) in the case of BECs confined
by the external potential $U(x)$ are modified as 
\begin{eqnarray}
u_{t} &=&\gamma u+i(\hbar /2m)u_{xx}+\left( i\sigma _{1}-\sigma _{2}\right)
|u|^{2}u+iv-i\hbar ^{-1}U(x)u,  \label{u2} \\
v_{t} &=&-\left( \Gamma +i\chi \right) v+i(\hbar /2m)v_{xx}+iu-i\hbar
^{-1}U(x)v.  \label{v2}
\end{eqnarray}
We have checked that, in the model introduced above, the addition of the
harmonic potential $U(x)=\left( K/2\right) (x-L/2)^{2}$, where $x=L/2$ is
the central point of the magnetic trap, gives rise to very persistent
periodic oscillations of the SP. Moreover, it can be verified that, in this
case, the SP behaves as a perfect quasi-particle, obeying the corresponding
equation of motion 
\begin{equation}
m\frac{d^{2}X_{p}}{dt^{2}}=-\frac{dU}{dx}\equiv -K(X_{p}-L/2),
\label{semiclassical}
\end{equation}
where $X_{p}$ is the coordinate of the pulse's peak (provided that
potential's strength $K$ is not too large). 

The simplest way to derive Eq. (\ref{semiclassical}) is to assume the
solution in the same boosted form as given by the Galilean transformation 
(\ref{k}), (\ref{c,omega}), but assuming that the velocity $c$ may be a
slowly varying function of time, i.e.,
\begin{eqnarray}
u &=&u_{0}\left( x-\int c(t)dt,t\right) \exp \left[ \frac{imc(t)}{\hbar}x-\frac{im}{2\hbar}\int c^{2}(t)dt\right] ,\;  \nonumber \\
v &=&v_{0}\left( x-\int c(t)dt,t\right) \exp \left[ \frac{imc(t)}{\hbar}x-
\frac{im}{2\hbar}\int c^{2}(t)dt\right] .  \label{generalized}
\end{eqnarray}
Then, the approximation (\ref{generalized}) should be substituted into the
balance equation for the net field momentum $P$, which in the presence of the external potential has the form
\begin{eqnarray}
\frac{dP}{dt}&\equiv &\frac{d}{dt}\int_{-\infty }^{+\infty }(-i\hbar)\left(
u^{\ast }u_x+v^{\ast }v_x\right) dx/\int_{-\infty}^{+\infty}(|u|^2+|v|^2)dx\nonumber\\
&=&\int_{-\infty
}^{+\infty }U(x)\frac{\partial }{\partial x}\left( |u|^{2}+|v|^{2}\right) dx/\int_{-\infty}^{+\infty}(|u|^2+|v|^2) dx  \label{dP/dt}
\end{eqnarray}
[note that this definition of the field momentum agrees with the relation 
$mc\equiv P$ in Eq. (\ref{E}) and $\int_{-\infty}^{+\infty}(|u|^2+|v|^2)dx$ is constant in time for Eq. (\ref{generalized})]. Finally, identifying $X_{p}\equiv \int c(t)dt$
and assuming that the potential $U(x)$ varies on a scale which is
essentially larger than the internal size of the pulse, one can easily
derive Eq. (\ref{semiclassical}) from Eq. (\ref{dP/dt}) by means of the
integration by parts.

Figure 5 displays the time dependence of $X_{p}$ in the harmonic potential 
$U(x)=0.005(x-L/2)^{2}$ for $m=0.1$. The initial peak position is 
$X_{p}(0)=L/4=15$, and the initial wavenumber is $k=0$ [see Eqs. (\ref{k})].
The time evolution of the peak position almost exactly follows the
corresponding solution $X_{p}(t)=30-15\cos (\sqrt{0.1}t)$ of Eq. (\ref
{semiclassical}). If two pulses are originally placed in the trap, they
perform persistent oscillations, periodically passing through each other in
the elastic fashion, like in Fig. 4.

Another physically relevant possibility (in the application to BECs) is to
consider a system with a narrow potential wall which separates a broad but
finite trap into two compartments. In other words, it is a double-well
configuration, which is frequently considered in the context of
one-dimensional BECs, but usually for the case of positive scattering length
(i.e., the self-repulsive nonlinearity), see, e.g., Ref. \cite{Luca}. To
simulate this situation, the potential $U(x)$ was taken (for instance) as 
\begin{equation}
U(x)=\left\{ 
\begin{array}{ll}
2, & \left| x-30\right| >15 \\ 
1, & \left| x-30\right| <0.5 \\ 
0, & 0.5<\left| x-30\right| <15
\end{array}
\right. \,.  \label{U}
\end{equation}
In this case, the initial SP was given a velocity corresponding to the
kinetic energy $E_{\mathrm{kin}}=0.8$ [see Eq. (\ref{E})], which is lower
than the central potential barrier in the expression (\ref{U}).

The result is displayed in Fig. 6, which demonstrates that the moving pulse
can tunnel across the central potential wall several times. Note that,
unlike the case displayed above for the case of the smooth harmonic
potential, in the present situation the pulse's motion does \emph{not} obey
the classical Newton's equation of motion, which may be explained by the
fact that the potential function (\ref{U}) varies steeply in space, breaking
the applicability condition for Eq. (\ref{semiclassical}). Further, we
notice that, in the course of the multiple tunnelings, the norm of the pulse
remains practically constant, while its kinetic energy gradually decreases.
Eventually, the tunneling ceases, and the SP finds itself trapped in one
compartment.

If the potential function is asymmetric, spatially periodic, and,
simultaneously, time-periodic, the quantum ratchet effect may be observed 
\cite{Magnasco}, \cite{Reimann}. To demonstrate this possibility, we adopt a
time-periodic sawtooth potential: 
\begin{equation}
U(x,t)=\left[ 1+0.8\cos (2\pi t)\right] \times \left\{ 
\begin{array}{ll}
7(10-x), & 9.5<x<10, \\ 
7(25-x), & 24.5<x<25, \\ 
7(40-x), & 39.5<x<40, \\ 
7(55-x), & 54.5<x<55, \\ 
0, & \mathrm{otherwise\,,}
\end{array}
\right. \,  \label{U2}
\end{equation}
so that the peak amplitude of the potential oscillates as $3.5\left[
1+0.8\cos (2\pi t)\right] $. We also assume periodic boundary conditions,
the spatial period being $15$.

Figure 7(a) displays the time evolution of the peak position for the SP with
the initial velocity $c_{0}=4$ [see Eq. (\ref{c,omega})]. As is seen from
the figure, the pulse steadily moves in the right direction. Note that the
initial kinetic energy of the pulse, $E_{\mathrm{kin}}=0.8$, is lower than
the average height, $\left\langle U_{\max }\right\rangle =3.5$, of the
potential peak, hence progressive motion is possible due to the tunnel
effect. The kinetic energy decreases as a result of the multiple tunnelings
as in Fig.~6; however, energy supply is possible in the time-periodic
potential. Thus, the nearly steady propagation is possible, as is seen in
Fig.~7(a). On the other hand, if the initial velocity is $c_{0}=-4$,
simulations demonstrate that, while the kinetic energy again decreases in
time, the energy supply is not efficient enough in this case to compensate
the loss, and the pulse gets trapped in a potential well after a transient,
see Fig. 7(b). Thus, the time-periodic sawtooth potential admits only the
unidirectional drift to the right, which conforms to the definition of
quantum rachets \cite{Reimann}.

\section{Conclusion}

In this work, we have proposed a model of a two-core system, based on an
equation of the Ginzburg-Landau (GL) type, coupled to another GL equation,
which may be linear or nonlinear. One core is active, being equipped with
linear gain, while the other one is lossy. The difference from previously
considered models is that the overall stabilization of the system is
provided not by the linear filtering (diffusion) term, but rather by cubic
or quintic dissipation in the active core. Physical realizations of the
model include several systems from nonlinear optics (semiconductor
waveguides or optical cavities), and a double-cigar-shaped BEC, in which one
``cigar'' is actually an atom laser. The replacement of the diffusion term
by the nonlinear loss is principally important, as diffusion is not possible
in these physical systems, while the nonlinear loss may occur. A stability
region for solitary pulses was identified in the relevant parameter plane by
means of numerical simulations. One border of the region can be predicted in
an analytical form by the perturbation theory. Moving pulses were considered
too, with the conclusion that collisions between them are completely elastic
(unless the relative velocity is too small), and they withstand multiple
tunneling through potential barriers without losing their coherence. The
existence of the robust quantum-rachet regime of motion for the pulses was
demonstrated as well.

\end{document}